\documentclass[pdflatex,sn-mathphys]{sn-jnl}

\jyear{2022}%

\theoremstyle{thmstyleone}%
\theoremstyle{thmstyletwo}%

\theoremstyle{thmstylethree}%

\usepackage{astjnlabbrev-nature} 

\usepackage{ulem}
\newcommand\arcmin{\mbox{$^\prime$}}%
\newcommand\arcsec{\mbox{$^{\prime\prime}$}}%

\newcommand{\Cii}{C\,{\sc ii}}

\newcommand{\Civ}{C\,{\sc iv}}
\newcommand{\Siiv}{Si\,{\sc iv}}

\graphicspath{{./}{./figs/}}

\begin{document}

\title[{[\Cii]}-emitters around a $z\approx5.7$ {\Civ} absorption system]{Compact [\Cii] emitters around a {\Civ} absorption complex at redshift 5.7}


\author*[1,2]{\fnm{Daichi} \sur{Kashino}}\email{kashino.daichi.v7@f.mail.nagoya-u.ac.jp}
\author[3]{\fnm{Simon J.} \sur{Lilly}}\email{simon.lilly@phys.ethz.ch}
\author[4]{\fnm{Robert A.} \sur{Simcoe}}\email{simcoe@space.mit.edu}
\author[5]{\fnm{Rongmon} \sur{Bordoloi}}\email{rbordol@ncsu.edu}
\author[3]{\fnm{Ruari} \sur{Mackenzie}}\email{mruari@ethz.ch}
\author[3]{\fnm{Jorryt} \sur{Matthee}}\email{mattheej@phys.ethz.ch}
\author[4]{\fnm{Anna-Christina} \sur{Eilers}}\email{eilers@mit.edu}

\affil*[1]{\orgdiv{Institute for Advanced Research}, \orgname{Nagoya University}, \orgaddress{\city{Nagoya}, \postcode{464-8601}, \state{Aichi}, \country{Japan}}}

\affil[2]{\orgdiv{Department of Physics}, \orgname{Nagoya University}, \orgaddress{\city{Nagoya}, \postcode{464-8602}, \state{Aichi}, \country{Japan}}}

\affil[3]{\orgdiv{Department of Physics}, \orgname{ETH Z{\"u}rich}, \orgaddress{\street{Wolfgang-Pauli-Strasse 27}, \city{Z{\"u}rich}, \postcode{8093}, \country{Switzerland}}}

\affil[4]{\orgname{MIT Kavli Institute for Astrophysics and Space Research}, \orgaddress{\street{77 Massachusetts Avenue}, \city{Cambridge}, \postcode{02139}, \state{Massachusetts}, \country{USA}}}

\affil[5]{\orgdiv{Department of Physics}, \orgname{North Carolina State University}, \orgaddress{\city{Raleigh}, \postcode{27695}, \state{North Carolina}, \country{USA}}}


\abstract{
The physical conditions of the circumgalactic medium are probed by intervening absorption-line systems in the spectrum of background quasi-stellar objects out to the epoch of cosmic reionization \citep{2006ApJ...637..648S,2007ApJS..171...29P,2007A&A...473..791F,2022Natur.606...59B}. A correlation between the ionization state of the absorbing gas and the nature of the nearby galaxies has been suggested by the sources detected either in Ly$\alpha$ or [\Cii]~158~$\mu$m near to respectively highly-ionized and neutral absorbers \citep{2021MNRAS.502.2645D,2021NatAs...5.1110W}.  This is also likely linked to the global changes in the incidence of absorption systems of different types and the process of cosmic reionization \citep{2006ApJ...640...69B,2011ApJ...743...21S,2017ApJ...850..188C,2018MNRAS.481.4940C,2019ApJ...882...77C,2019ApJ...883..163B}.
Here we report the detection of two [\Cii]-emitting galaxies at redshift $z \approx 5.7$ that are associated with a complex high-ionization {\Civ} absorption system.
These objects are part of an overdensity of galaxies and have compact sizes ($<2.4$~kpc) and narrow line widths ($\mathrm{FWHM}\approx62\textrm{--}64~\mathrm{km~s^{-1}}$).
Hydrodynamic simulations predict that similar narrow [\Cii] emission may arise from the heating of small ($\lesssim 3~\mathrm{kpc}$) clumps of cold neutral medium or a compact photodissociation region \citep{2013MNRAS.433.1567V,2015ApJ...813...36V}. The lack of counterparts in the rest-frame ultraviolet indicates severe obscuration of the sources that are exciting the [\Cii] emission.
These results may suggest a connection between the properties of the [\Cii] emission, the rare overdensity of galaxies and the unusual high ionization state of the gas in this region.
}

\keywords{Metal absorption systems, absorber host galaxies, intergalactic medium}


\maketitle

\section*{Main}

Two line emitters are detected within a mosaiced submilimeter observation with Atacama Large Millimeter Array (ALMA) that covers a field of about 23$\arcsec$ (138~physical kpc at $z=5.7$) in radius centered on the position of QSO J1030$+$0524 at redshift $z_\mathrm{QSO}=6.308$.  We interpret these two emission lines to be [\Cii]~$158~\mu\mathrm{m}$ at redshifts close to a complex {\Civ} absorption system that is found with three distinct absorbers at $z_\mathrm{abs}=5.7246$, 5.7411, and 5.7443 in a deep high-resolution spectrum of the QSO.  
These two sources were discovered in blind line searches within a larger datacube that covers a circular survey area of $23\arcsec$ (138~physical~kpc (pkpc) at $z=5.7$) and a total velocity interval $7380~\mathrm{km~s^{-1}}$, corresponding to $\approx 76~$comoving~Mpc (cMpc). No other significant lines were detected.  We estimated the reliability of these detections both to be $>95\%$.  The probability for these two line sources to be unrelated CO interlopers at low redshifts was calculated to be 1.5\%, and this possibility is further disfavored by other ancillary data in this region (see Methods).

\begin{figure}[t]
\centering
\includegraphics[width=4.7in]{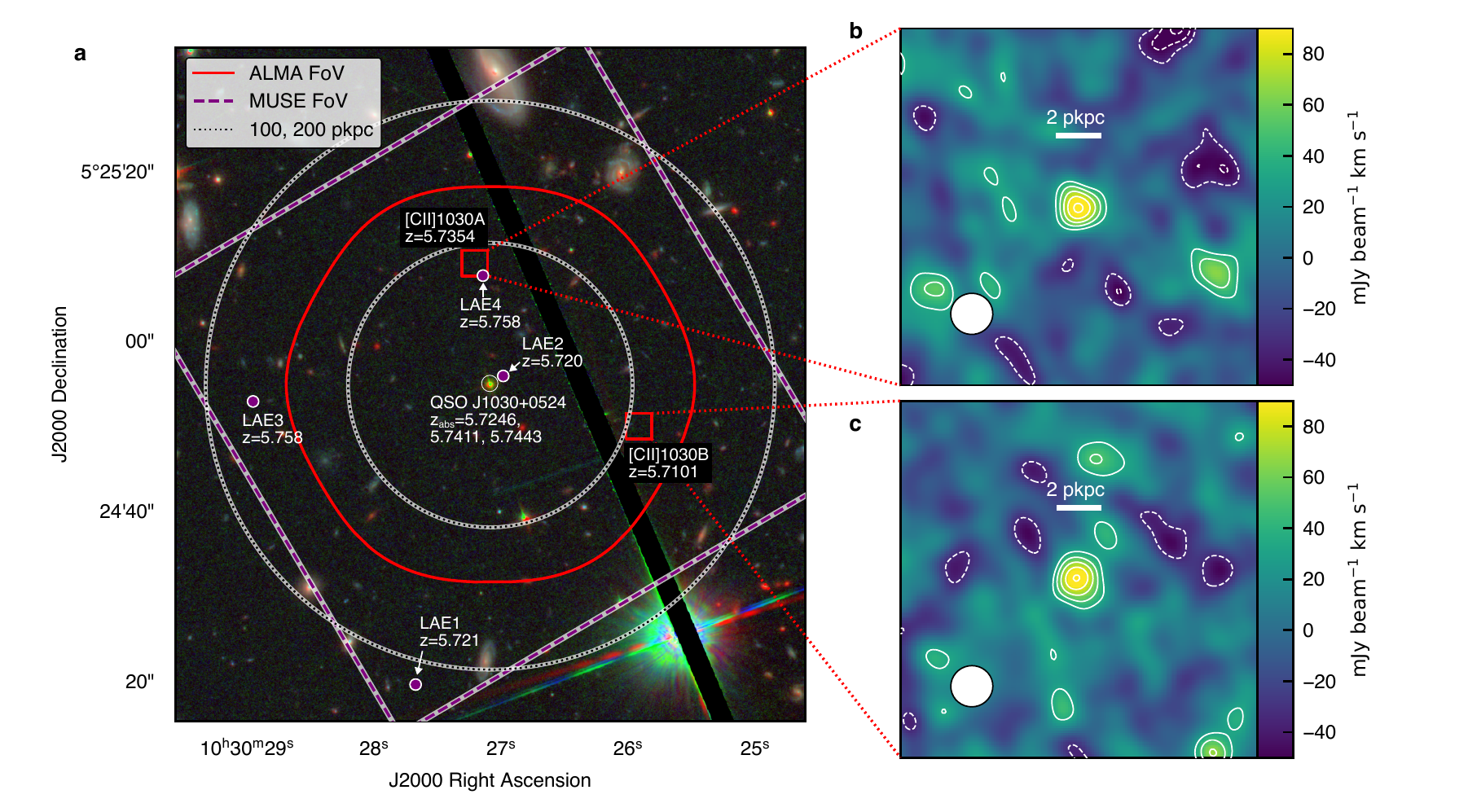}
\caption{
\textbf{ALMA observations of the two [\Cii] sources.} \textbf{a}, the field of QSO J1030$+$0524.  The background is a composite color image of the field with the Hubble Space Telescope images in the F775W, F850LP, andf F160W bands.  The field-of-view of the ALMA observation is marked by the red line. 
The [\Cii]-detected sources ([\Cii]1030A and [\Cii]1030B) are marked by red squares.  The large circles of the dotted lines denote the radii of 100 and 200 physical kpc from the central QSO.  For comparison, the purple dots indicate the locations of four Ly$\alpha$-emitters (LAEs \#1--4) found in the MUSE datacube whose field-of-view is marked by the large square of the purple dashed line \citep{2021MNRAS.502.2645D}.  \textbf{b} and \textbf{c}, the [\Cii] moment-0 maps ($3\arcsec \times 3\arcsec$) of [\Cii]1030A and [\Cii]1030B, respectively.  The solid (dashed) contours mark positive (negative) steps of $1\sigma$ ($\approx0.25~\mathrm{mJy~km~s^{-1}}$ per beam) starting at $2\sigma$ ($-2\sigma$).  The sizes of the synthesized beams are indicated at the bottom left.  
\label{fig:main}}
\end{figure}

\begin{figure}[t]
\centering
\includegraphics[width=3.4in]{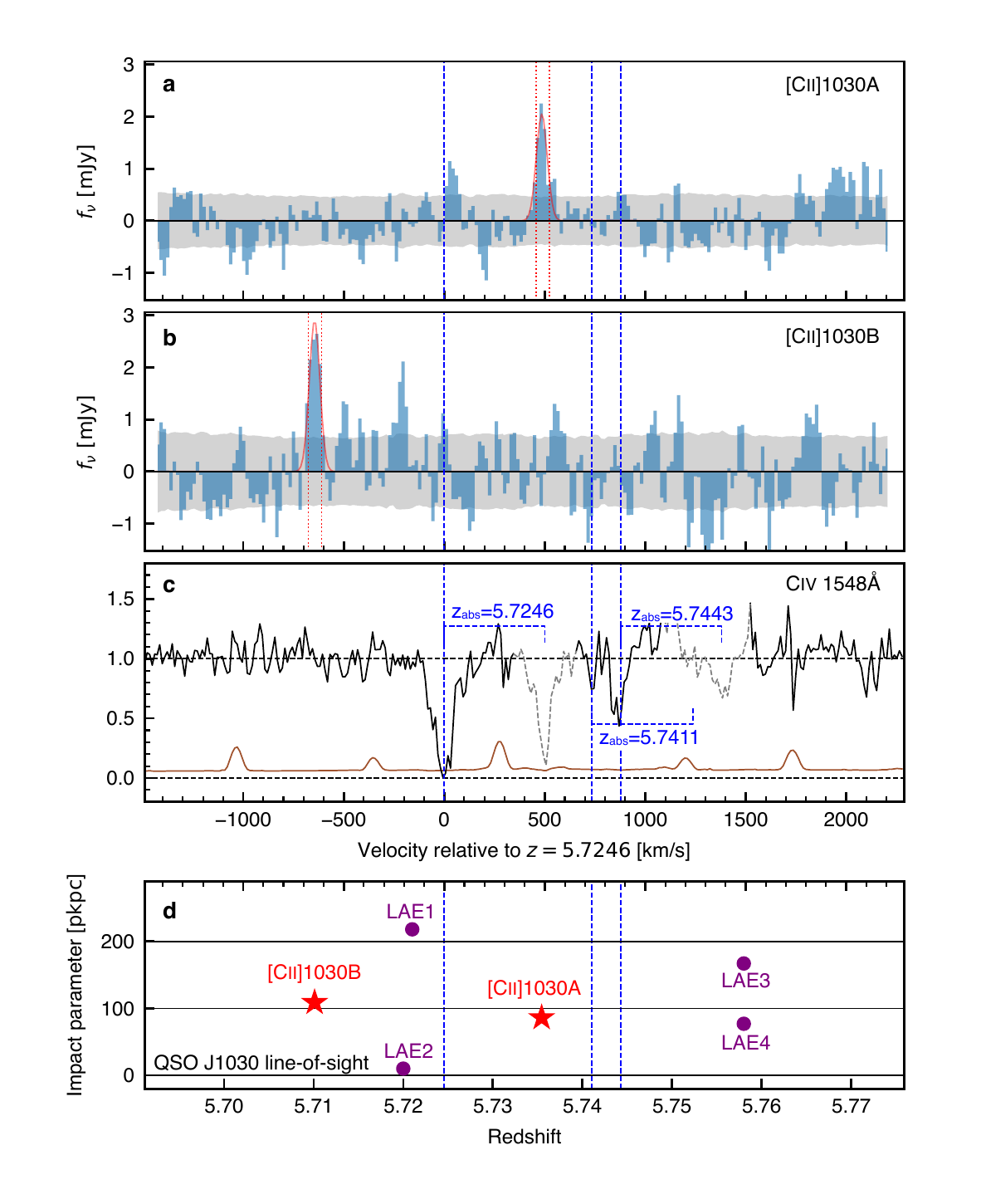}
\caption{
\textbf{Emission and absorption spectra from the [\Cii] sources and QSO J1030$+$0524.} \textbf{a} and \textbf{b}, the ALMA spectra of the [\Cii]~$158~\mathrm{\mu m}$ emission of the [\Cii]1030A and B, respectively.  The gray shaded region indicates the $1\sigma$ noise.  The red solid line shows a single-Gaussian model fit to the data. The red dotted vertical lines mark the channel intervals that were collapsed to generate the [\Cii] moment-0 maps.  The blue dashed vertical lines indicate the location of three absorbers.  The origin of the velocity scale was chosen to correspond to the strongest {\Civ} absorption system at $z_\mathrm{abs}=5.7246$.
\textbf{c}, the absorption profiles of the {\Civ} doublet in the continuum-normalized spectrum of the QSO J1030$+$0524.  
The {\Civ} doublet lines in the same system are connected.  The velocity scale is applicable to the bluer line (rest-frame 1548~{\AA}) of the doublet.  The gray-dashed regions in the spectrum indicate the redder 1550~{\AA}) line.  The brown line indicates the 1$\sigma$ noise spectrum.
\textbf{d}, impact parameter versus redshift and radial velocity relative to the strongest {\Civ} absorber at $z_\mathrm{abs}=5.7246$.  The $y=0$ line corresponds to the QSO sightline.  The [\Cii] sources are highlighted by red stars and four MUSE LAEs are presented by purple dots.
\label{fig:main_spec}}
\end{figure}

The first source, referred to as [\Cii]1030A, at $z_\textrm{[C\,{\sc ii}]}=5.7354$, has an impact parameter to the quasar sightline\footnote{The transverse separation from the QSO line of sight at the redshift of the intervening system.} of 86~pkpc.  The second source, [\Cii]1030B, at $z_\textrm{[C\,{\sc ii}]}=5.7101$ is at $109~\mathrm{pkpc}$ (Fig~\ref{fig:main}a).  These sources have the maximum peak signal-to-noise ratios (S/N) of $>6.0$ when collapsing the emission line channels over $66~\mathrm{km~s^{-1}}$ (Methods).  The [\Cii] moment-0 intensity maps of these [\Cii] sources are shown in Fig.~\ref{fig:main}b and c.  Interestingly, these two sources are not spatially resolved under the synthesized FWHM beam size of $0.35\arcsec$, putting  a strong constraint on the FWHM size of the [\Cii] sources to be less than 2.4~pkpc.  The spectra of these sources are shown in Figure \ref{fig:main_spec}a,b, in comparison with the {\Civ} absorption profiles in the continuum-normalized spectrum of the QSO J1030$+$0524 (Fig.~\ref{fig:main_spec}c; Methods). 
These ALMA spectra are extracted within radius $0.24\arcsec$ of the peak position.  By fitting a simple Gaussian profile, we estimate the FWHM line widths to be $64\pm11~\mathrm{km~s^{-1}}$ ([\Cii]1030A) and $62\pm11~\mathrm{km~s^{-1}}$ (B), and the velocity-integrated [\Cii] flux densities to be $0.174\pm0.027~\mathrm{Jy~km~s^{-1}}$ (A) and $0.276\pm0.044~\mathrm{Jy~km~s^{-1}}$ (B) after correcting for the primary beam response (Methods).  The two dotted lines represent the four spectral channels, equivalent to a velocity width of $66~\mathrm{km~s^{-1}}$, used to create the moment-0 maps.  The total [\Cii] luminosity is $L_\textrm{[C\,{\sc ii}]} = (1.61\pm0.24)\times10^8~{L_\odot}$ and $(2.21\pm0.36)\times10^8~{L_\odot}$, respectively, for [\Cii]1030A and B.  
Figure~\ref{fig:main_spec}d summarizes the kinematic structure of the [\Cii] sources relative to the absorption systems.
The [\Cii]1030A ($z_\textrm{[C\,{\sc ii}]}=5.7354$) has a velocity difference of $\Delta V = +485~\mathrm{km~s^{-1}}$ relative to the stronger absorption system at 
$z_\mathrm{abs}= 5.7246$.  The [\Cii]1030B ($z_\textrm{[C\,{\sc ii}]}=5.7101$) is at $\Delta V =-645~\mathrm{km~s^{-1}}$.
Note that [\Cii]1030A has a velocity difference of $\Delta V=-250~\mathrm{km~s^{-1}}$ relative to its closest (and weakest) absorption system at $z_\mathrm{abs}=5.7411$.
The ALMA measurements (primary beam corrected values) of the [\Cii] sources are summarized in Table \ref{tb:meas}.

The impact parameters of these [\Cii] sources are within the range seen previously for associations between metal absorbers and (non [\Cii]) galaxies at lower redshifts \citep{2006ApJ...637..648S,2021MNRAS.508.4573D}.
The velocity offsets are also comparable to the velocity dispersion of proto-clusters identified at similar high redshifts \citep{2020ApJ...888...89T} and for companion [\Cii] sources ($\lesssim 100~\mathrm{pkpc}$) observed around $z>6$ luminous quasars \citep{2020ApJ...904..130V}.  A similar extent is also found in zoom-in simulations that follow the formation of a massive proto-cluster at $z \sim 6$ \citep{2022MNRAS.513.2118Z}.  All these arguments support a physical connection between the [\Cii] sources and the {\Civ} absorbing gas through a common membership of some larger structure(s) that may enclose all these systems, rather than the {\Civ} absorption coming from gas falling into or flowing out of these individual sources.

This statement is further strengthened by the locations of the four Lyman-$\alpha$ emitters (LAEs) that have been discovered in this region.
Figures~\ref{fig:main}a and \ref{fig:main_spec}d mark these LAEs (purple circles) detected within $\pm1500~\mathrm{km~s^{-1}}$ of the {\Civ} absorbers with the Multi-Unit Spectroscopic Explore (MUSE) on the Very Large Telescope (VLT) \citep{2021MNRAS.502.2645D}.
The two [\Cii] sources and the four LAEs are all quite distinct sources.
Although one of the LAE (\#4; $z=5.758$) has a small separation of only $1.7\arcsec$ from [\Cii]1030A (equivalent to $10~\mathrm{pkpc}$), the redshift difference between these two sources corresponds to more than $1000~\mathrm{km~s^{-1}}$ suggesting that they are different objects.
The typical impact parameter of the galaxies found here,  $\sim100~\mathrm{pkpc}$, is comparable to the possible most massive halos of mass $10^{13}~M_\odot$ at this epoch.
However, the radial velocity dispersion of all six sources, $\sigma(v_\mathrm{r})=830\pm270~\mathrm{km~s^{-1}}$, is larger than expected for such halos $(\sim200~\mathrm{km~s^{-1}})$.
Therefore, the system is unlikely to be virialized.  
The discovery of the [\Cii] sources has further strengthened this hypothesis.

These two sources are both characterized by compact sizes (FWHM diameter $<2.4$~pkpc) and narrow [\Cii] emission line profiles ($64\pm11~\mathrm{km~s^{-1}}$ and $62\pm11~\mathrm{km~s^{-1}}$ respectively for [\Cii]1030A and B).  
These properties differ from other high-redshift [\Cii] emitters reported before, which tend to have a major axis FWHM of $>4~\mathrm{pkpc}$ and a [\Cii] line width FWHM of $>200~\mathrm{km~s^{-1}}$ \citep{2016ApJ...833...71A,2017Sci...355.1285N,2020ApJ...900....1F,2022ApJ...934..144F}.
Hydrodynamic simulations have suggested that the [\Cii] emission can be dominated by a narrow (FWHM~$<100~\mathrm{km~s^{-1}}$) line peoduced by molecular clouds and photodissociation regions within a compact region ($\sim2~\mathrm{pkpc}$) of a galaxy \citep{2015ApJ...813...36V,2015ApJ...814...76O}, or by small ($\sim2~\mathrm{pkpc}$) clumps of cold neutral medium (CNM) that displaced  by several pkpc from the main body of a galaxy undergoing intense star formation \citep{2013MNRAS.433.1567V}.  
Note that the contrast of emission from CNM clumps may be substantially reduced as the cosmic microwave background (CMB) radiation increases in temperature with redshift (the CMB is at 18.5~K at $z=5.7$) so as to be similar to the excitation temperature of the emission \citep{2015ApJ...813...36V}.

In both cases, the [\Cii] emission is excited by FUV radiation from ongoing star formation.  
Using an empirical relation between $L_\textrm{[C\,{\sc ii}]}$ and SFR, the measured $L_\textrm{[C\,{\sc ii}]}$ would correspond to a [\Cii]-based SFR of $\mathrm{SFR_\textrm{[C\,{\sc ii}]}} = 14~M_\odot~\mathrm{yr^{-1}}$ ([\Cii]1030A) and $19~M_\odot~\mathrm{yr^{-1}}$ (B), respectively, with an expected uncertainty of a factor of 2 due to the scatter \citep{2014A&A...568A..62D,2020A&A...643A...3S}.  Theoretical studies predict consistent values for gas clouds of solar metallicity \citep{2015ApJ...813...36V}.  
These sources were not detected in a rest-frame FIR continuum, yielding only a loose $2\sigma$ upper limit to the IR-based SFR of $\approx30~M_\odot~\mathrm{yr^{-1}}$, which is still consistent with the [\Cii]-inferred SFR.

For our claimed [\Cii] sources, no significant counterparts have so far been found either in the Ly$\alpha$ emission line or in the rest-frame FUV continuum, despite the availability of deep Hubble Space Telescope (HST) images and the MUSE datacube.  The lack of detection in an HST image ($\approx 9000${\AA} observed or $\approx1350${\AA} in the rest-frame) gives a $2\sigma$ upper limit of $\sim 59~\mathrm{nJy}$, which corresponds at this redshift to a FUV-based SFR of $\sim2~M_\odot~\mathrm{yr}^{-1}$ (methods), considerably lower than the [\Cii]-based SFRs. This therefore suggests that the FUV emission from SFR at the required levels must have been severely attenuated by dust, which would also explain the absence of Ly$\alpha$.

The absence of detectable FUV emission is also in some tension with the ``normal'' [\Cii] sources that have been detected at similar redshifts, which are usually detected in rest-frame UV and Ly$\alpha$ emission \citep{2018MNRAS.478.1170C,2020A&A...643A...1L}, with typical UV obscuration fractions $\sim40\%$ \citep{2020A&A...643A...4F}.  A few other reports of narrow ($\mathrm{FWHM} =50\textrm{--}90~\mathrm{km~s^{-1}}$) [\Cii] sources at $z>6$ have come from following up sources that had already been detected in Ly$\alpha$ and the FUV \citep{2016ApJ...829L..11P,2017ApJ...836L...2B}.This difference with the sources claimed here may be partially explained by the difference in observing strategy: our own study was based on a blind search as against the follow-up observations of previously detected FUV sources.
However, heavy obscuration may be at odds with the unusual spatial compactness and narrow spectral width of the [\Cii] sources, which suggest they are unlikely to be mature massive systems, which are observed as FIR-bright dusty star-forming galaxies at lower redshifts.  Rather, they may represent a different population of galaxies specific to the high-redshift universe.
On the other hand, simulations show that selective attenuation by dense dust clouds that embed young stellar clusters may lead to absorption of most of the UV luminosity \citep{2018MNRAS.477..552B}.

We cannot therefore yet make a conclusive statement about the nature of these [\Cii]-emitting sources.  
The threefold rarity of simultaneously having 1) these unusual [\Cii] sources, 2) a strong overdensity of galaxies, and 3) the rich high ionization {\Civ} absorption system (rare at these redshifts), however, suggests that there may be some connection between the observed [\Cii] properties and the unusual high-ionization conditions that are seen in this particular region, which we are observing at the time that the cosmic reionization of the Universe was approaching completion.  Addressing this question will require us to bridge between the rest-frame FUV and FIR views of the high redshift universe in order to unveil the stellar content and the star formation at wavelengths that are much less affected by dust.  Deep NIR observations with JWST will soon enable us to clarify the origin of these unusual [\Cii] lines and the links between different physical processes occurring within this extreme environment during the end stages of cosmic reionization.


\clearpage
\vspace{10cm}

\clearpage
\begin{table}[h]
\begin{center}
\begin{minipage}{\textwidth}
\caption{\textbf{ALMA measurements for the two [\Cii] sources.}  The impact parameter is the transverse separation from the QSO J1030$+$0524 sightline in physical kpc, $\Delta V$ is the velocity shift relative to each of the three descrete {\Civ} absorption components.  The [\Cii] flux and luminosity are corrected for the primary beam response.}\label{tb:meas}%
\begin{tabular}{@{}lcc@{}}
\toprule
& [\Cii]1030A & [\Cii]1030B \\
\midrule
J2000 Right Ascension & $10^\mathrm{h}30^\mathrm{m}27.215^\mathrm{s}$ & $10^\mathrm{h}30^\mathrm{m}25.921^\mathrm{s}$ \\
J2000 Declination     & $+05^\mathrm{d}25^\mathrm{m}09.36^\mathrm{s}$ & $+05^\mathrm{d}24^\mathrm{m}50.20^\mathrm{s}$ \\
Impact parameter (pkpc)  & 86       & 109  \\
Redshift                 & 5.7354   & 5.7101  \\
$\Delta V$ relative to $z_\mathrm{abs}=5.7246$ ($\mathrm{km~s^{-1}}$) & $+485$ & $-645$  \\
$\Delta V$ relative to $z_\mathrm{abs}=5.7443$ ($\mathrm{km~s^{-1}}$) & $-392$ & $-1518$  \\
$\Delta V$ relative to $z_\mathrm{abs}=5.7411$ ($\mathrm{km~s^{-1}}$) & $-250$ & $-1377$  \\
FWHM spatial size ($\mathrm{pkpc}$) & $<2.4$ & $<2.4$ \\
{[\Cii]} flux ($\mathrm{Jy~km~s^{-1}}$) & $0.174 \pm 0.027$ & $0.276 \pm 0.044$ \\
FWHM line width ($\mathrm{km~s^{-1}}$) & $64 \pm 11$ & $62 \pm 11$ \\
{[\Cii]} luminosity ($10^8~L_\odot$) & $1.61\pm0.24$ & $2.21\pm0.36$ \\
\botrule
\end{tabular}
\end{minipage}
\end{center}
\end{table}

\clearpage

\clearpage
\section*{Methods}

\setcounter{figure}{0}
\setcounter{table}{0}
\renewcommand{\figurename}{Extended Data Fig.}
\renewcommand{\tablename}{Extended Data Table}

\subsection*{Cosmology}
Throughout this paper, we adopted a flat $\Lambda$ Cold Dark Matter cosmology with $\Omega_\mathrm{\Lambda}=0.69$, $\Omega_\mathrm{M}=0.31$ and  $H_0=67.7~\mathrm{km~s^{-1}~Mpc^{-1}}$ \citep{2020A&A...641A...6P} when calculating physical parameters.

\subsection*{ALMA observations and data reduction}

The ALMA observations reported here were carried out in 2018 September (program ID: 2017.1.000621.S) with the array in a configuration of 43 12~m antennas with baselines from 15--1397~m.  The primary beam of the 12~m ALMA antennas is $\approx 45\arcsec$ in diameter at $\sim 260~\mathrm{GHz}$. 
The dataset used in this study covers a mosaic of seven offset but overlapping pointings that together cover a roughly circular field of $\approx 0.46~\mathrm{arcmin}^2$ centered on the optical coordinate of the background QSO J1030$+$0524 ($10^\mathrm{h}30^\mathrm{m}27^\mathrm{s}.092$, $+05^\circ24\arcmin55.02\arcsec$; \cite{2015ApJS..219...12A}), with 41 min of total on-source integration.
Extended Data Fig.~\ref{fig:mosaic} shows the mosaiced field coverage, which is approximately a circular region of $23\arcsec$ in radius, and the resulting antenna response, which is 67\% and 47\%, respectively, at the positions of [\Cii]1030A and [\Cii]1030B.

The datacubes are generated from four spectral windows (SPWs), each of which has a band width of $\approx1.875$~GHz.  Of these, two SPWs are adjusted to cover the three absorbers at $z_\mathrm{abs}=5.7246, 5.7411$ and $5.7443$ in the [\Cii]~$158~\mathrm{\mu m}$ emission line, respectively, with a minimum overlap at $282.3~\mathrm{GHz}$ to yield a continuous frequency coverage of 280.5--284.0~GHz.  The other two SPWs cover other separate frequency ranges around  $294~\mathrm{GHz}$, where there is no corresponding absorption system seen in the quasar spectrum.  
Each SPW was divided into 128 channels, each with a width of 15.6~MHz (16.6~km~s$^{-1}$), and the instrumental velocity dispersion corresponds to 2 channels ($33~\mathrm{km~s^{-1}}$).
 
The ALMA data was processed using the CASA \citep{2007ASPC..376..127M} pipeline for ALMA (version 5.4), using the standard calibration procedure.  The final datacubes and the continuum images were obtained using CASA \texttt{tclean} task with natural weighting to maximize sensitivity.  In doing so, we adopted uvtaper (a Gaussian taper on the weights of the $uv$ data) of $0 \times 220~\mathrm{mas}$ and a position angle (PA) $119~\mathrm{degree}$ in order to circularize the synthesized beam as much as possible.  The resulting beam is $0.35\arcsec$ in FWHM diameter.
The continuum image that collapses all the four SPWs has a $1\sigma$ sensitivity of $31~\mathrm{\mu Jy}$ per beam.  For the moment-0 [\Cii] maps, the four channels (62.5~MHz; $\approx 66~\mathrm{km~s^{-1}}$) that cover the emission line were collapsed resulting in a $1\sigma$ sensitivity of $4.1~(4.2)~\mathrm{mJy~km~s^{-1}}$ per beam at the observed frequency of the [\Cii]1030A ([\Cii]1030B).

\subsection*{Blind line search}

We employed the \texttt{findclumps} algorithm \citep{2016ApJ...833...67W}, implemented in the \texttt{interferopy} package for Python \citep{2021zndo...5775604B}, for our blind search for emission-line candidates.  Briefly, \texttt{findclumps} performs a top-hat convolution to a datacube to generate a set of collapsed images, and searches for peaks exceeding a certain S/N threshold.  Clumps that are detected with small offsets in spatial position and frequency are cropped as duplicates, keeping the one with the highest S/N in the final catalog.
We ran the line search with frequency window sizes from 4 to 17 channels (corresponding to $\approx 66 \textrm{--} 282~\mathrm{km~s^{-1}}$) and adopted the offset thresholds of $1\arcsec$ and $280~\mathrm{MHz}$ ($300~\mathrm{km~s^{-1}}$) for cropping duplicates.

\begin{figure*}[t]
\centering
\includegraphics[width=4.32in]{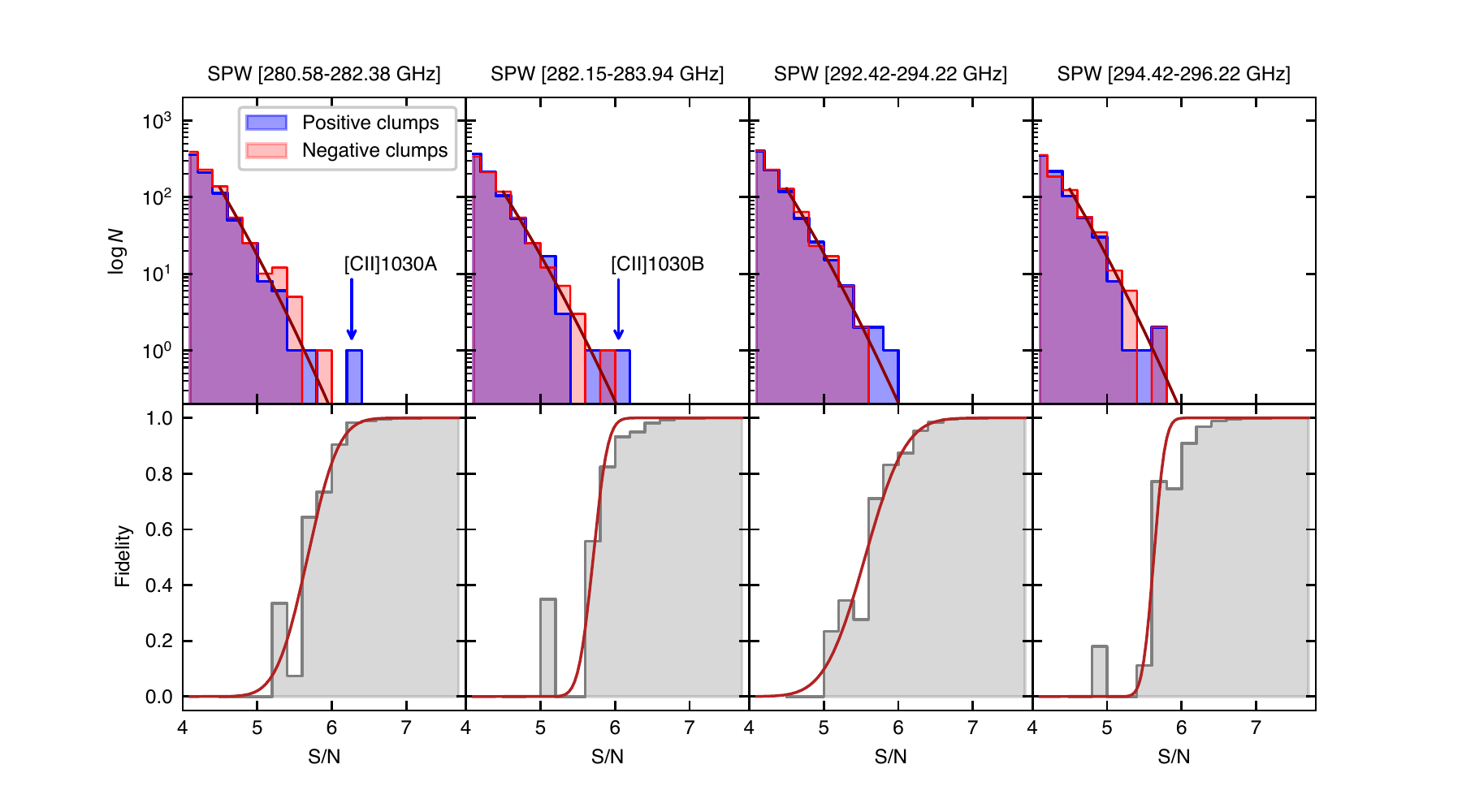}
\caption{
\textbf{Fidelity in our line search.}
Upper panels: number of positive (blue) and negative (i.e., noise; red) peaks detected in the four ALMA datacubes (one column for each).  The frequency range of each SPW is denoted at the top.  Lower panels: the fidelity at given S/Ns with the model fit as an error function (solid red line).
\label{fig:fidelity}
}
\end{figure*}

To assess the significance of the detections, we performed the same line search for (unphysical) negative peaks and computed the fidelity of the detections, which represents the probability that a positive line detection at a given S/N is real, following the methods in the literature \citep{2016ApJ...833...67W}.
Extended Data Fig.~\ref{fig:fidelity} shows the number of positive and negative peak detections (upper panels) and the fidelity (lower panel) as a function of S/N.  Expressing the fidelity with an error function, we found that 95\% fidelity is reached as $\mathrm{S/N}=5.8\textrm{--}6.2$, with only small variations between the SPWs. 
We selected the two claimed [\Cii] sources ([\Cii]~1030A and B), detected in four-channel collapsed images of two SPWs, as the only sources that have $> 6\sigma$ significance and found that there are no spurious negative signals at this $6.0\sigma$ level of significance.  The fidelity of the two [\Cii] sources were computed to be 96\% (A) and 99\% (B) respectively.

For each of the three absorption systems, a velocity interval of $\pm 700~\mathrm{km~s^{-1}}$ was taken as the relevant velocity range within which a source could be considered to be associated with the absorber.  This yield a contiguous velocity interval of 2280~$\mathrm{km~s^{-1}}$, from $-700~\mathrm{km~s^{-1}}$ of the lower-$z$ absorber's redshift ($z_\mathrm{abs}=5.7246$) to $+700~\mathrm{km~s^{-1}}$ of the higher-$z$ absorber's redshift ($z_\mathrm{abs}=5.7443$), which cover all of the the two [\Cii] sources and four LAEs detected in the MUSE datacube.  The small chance that the two most significant peaks, if actually spurious, both lie within this physically significant region of association, boosts the fidelity values of the claimed [\Cii] detections to $>99.9\%$.

\subsection*{Reliability check of the detections}

\begin{figure}[t]
\centering
\includegraphics[width=4.0in]{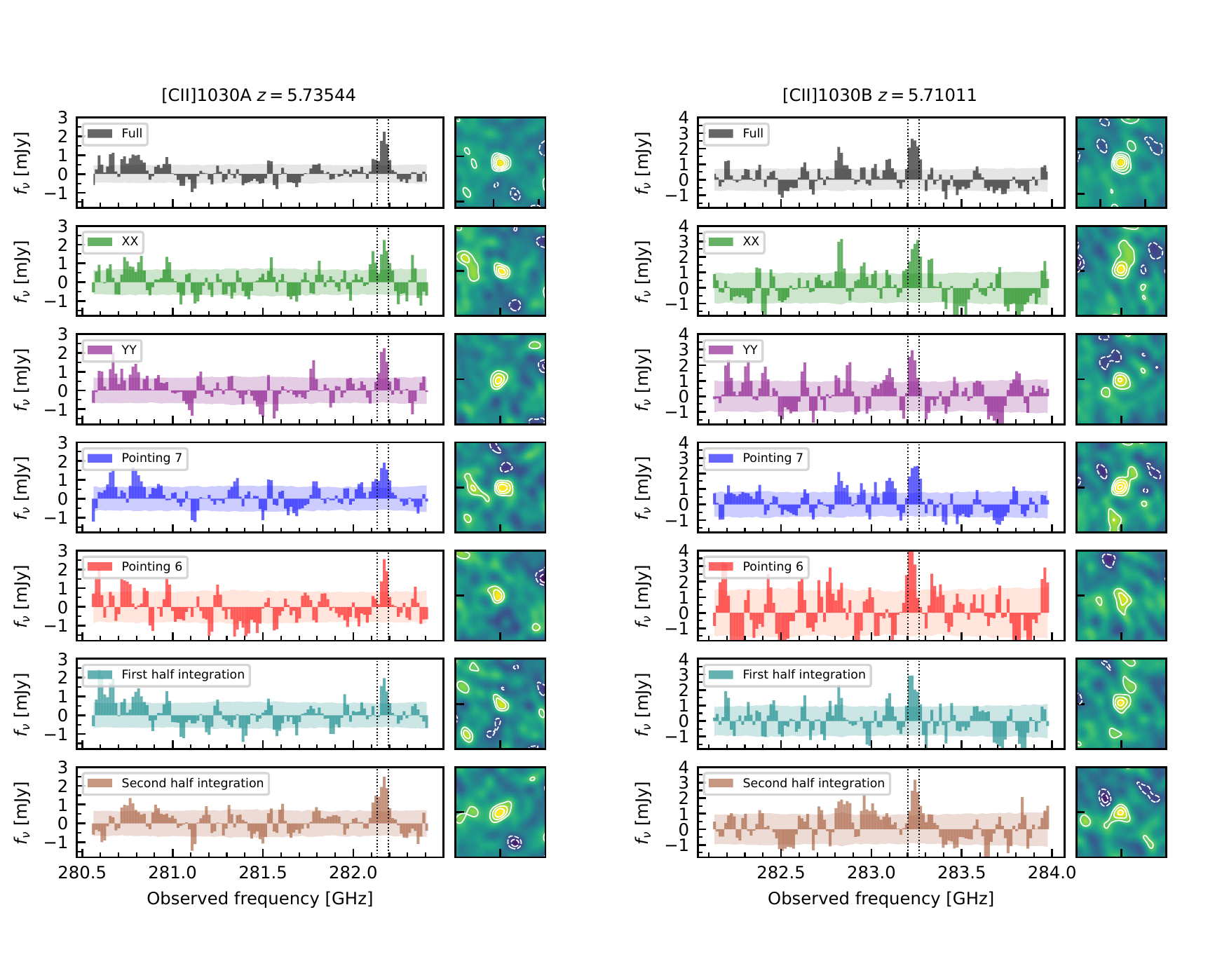}
\caption{\label{fig:reliability}
\textbf{
Reliability check of the [\Cii] detection.}  For each of [\Cii]1030A (left) and B (right), the spectra and the [\Cii] moment-0 maps ($2\arcsec \times 2\arcsec$) constructed in the reliability tests are compared to those from the full dataset that are shown in the top row.  
The second and third rows show the results under only XX or YY polarization.  The forth and fifth rows show the results for the first and second half exposure time intervals.  The sixth and seventh rows show the results from processing each of the two individual ALMA pointings that cover the source position.  In the spectral plots, the collapsed channel window is shown by dotted vertical lines.  In the [\Cii] intensity maps, the solid (dashed) contours mark positive (negative) steps of $1\sigma$ r.m.s starting at $2\sigma$ ($-2\sigma$).  All of these products from partial dataset exhibit significant signals that coincident both spatially and spectrally to those seen in the full data.}
\end{figure}

We further examined the reliability of the two claimed sources by analysing independent sub-sets of the data.  We processed separately the two independent XX and YY polarization correlations.  The [\Cii] intensity maps were generated from the same emission line channels, and the one-dimensional spectra were extracted from the same spaxels as the full analysis.  As shown in Extended Data Fig.~\ref{fig:reliability}, the line emission is always detected with consistent observed fluxes at $4.2\textrm{--}4.7\sigma$ at the expected spatial position (within $<0.2~\arcsec$) and at the expected frequency in each of the two independent data sets. 

\begin{figure}[t]
\centering
\includegraphics[width=3in]{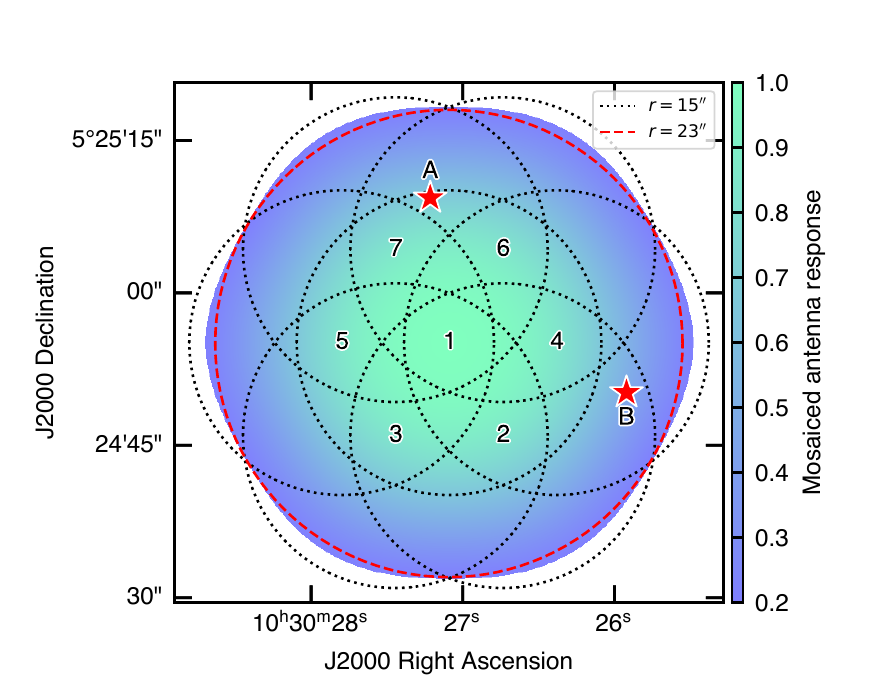}
\caption{
\textbf{Antenna response (primary beam) map of the mosaic ALMA observation.}  The response is normalized to one at the field center and survey field is defined as the area of $\approx23\arcsec$ in radius where the relative response is above 20\%.
The primary beam size of the individual 7 pointings ($r=15\arcsec$) that compose the mosaic observation is shown by dotted-line circle with the pointing number from 1 to 7.  The positions of the two [\Cii] sources ([\Cii]1030A and B) are marked by red stars.
\label{fig:mosaic}}
\end{figure}

Since the ALMA observations were in the form of a mosaic of seven offset but overlapping pointings, both the claimed sources in fact appear (at a different location relative to the pointing center) in two independent pointings (Figure \ref{fig:mosaic}), and these pointings could therefore be reduced separately.  The moment-0 maps and the spectra, created in the same way as above, are shown in Extended Data Fig.~\ref{fig:reliability}.  The claimed sources were detected with consistent fluxes at $3.4\textrm{--}5.3\sigma$. 

Finally, a similar test was also carried out using a temporal split of the data stream into two equal segments.
These two datasets may not strictly be independent because the two observations exposures were made serially without changes in the instrumental setups, but serves as a check whether the signals come from random noise or not: it should be extremely rare that spurious sources due to random noise are detected consistently in two different exposures.
Again, the line emission is consistently detected at the same spatial positions and at the expected frequencies in both the first and second halves of the data, at signal to noise ratios of between $3.9\textrm{--}4.9\sigma$.

These tests all convincingly indicate that the detected lines are plausibly real signals from astronomical sources and not random noise or other spurious signals.

\subsection*{Line identification}

We now address the question of whether the line emission is indeed the [\Cii]~158~$\mu$m emission line at the $z\approx5.7$ redshift of the {\Civ} absorption system or some other transition at lower redshift produced by unrelated foreground objects (background sources can be safely neglected).

The most likely low-redshift interlopers would be CO(3--2) (at $z\approx0.22$) or higher-$J$ CO emission ($z\approx 0.63 \textrm{--} 2.3$).  Using published CO luminosity functions for the redshift range between 0 and 6 \citep{2016MNRAS.461...93P}, we estimated that the expected number of CO sources within the whole datacube should be approximately 0.4 sources.
In doing so, we considered CO transitions up to $J=8\textrm{--}7$ and used the published Schechter luminosity function parameters derived for each CO transition at the closest redshift. The luminosity function of CO(6--5) at $z=2$ was adopted also for CO(7--6) and CO(8--7) as no measurement is provided for these highest transitions.  We then considered the expected number of CO sources that should be detected, at a line flux brighter or equal to the observed flux of the (fainter) [{\Cii}]1030A, and within the circular survey volume of radius 23$\arcsec$ and the frequency (redshift) interval corresponding to the whole ALMA datacube composed of four SPWs.  The results are summarized in Extended Data Table \ref{tb:CO_interlopers}.

The chance of finding two unrelated CO sources within the velocity range of interest (across $2280~\mathrm{km~s^{-1}}$ around the absorption systems; see above) is therefore of order $(0.4 \times 2280/7380)^2\approx 1.5\%$, which we consider to be very unlikely.
Even if we ignore the expected number of CO sources and consider only our observed number of two detected sources within the datacube, the chance that two unrelated emission lines are both found within the limited velocity range of interest is still 
$(2280/7380)^2\approx9.5\%$.
These probabilities therefore suggest that the two emission lines are indeed [\Cii] emission associated to this redshift range.

Non-detection of both sources in available deep HST images at $\sim9000$~{\AA} (see below) also argues against any interpretation that places these sources at low redshift ($z\lesssim2$). This is because, in that case, their stellar continuum should be detectable.  The non-detection is consistent with their lying at high redshift ($z \gtrsim 5$) if their FUV and Lyman-$\alpha$ emission is intrinsically faint or obscured by dust.  We therefore identify these two emission line signals as [\Cii] emission at $z\approx5.7$.

\subsection*{The properties of the [\Cii] sources}

We measured the line properties by fitting a simple Gaussian profile to the observed one-dimensional spectra extracted from the spaxels enclosed within $0.24\arcsec$ of the peak position of the [\Cii] sources.   
The velocity-integrated [\Cii] line flux is measured to be $0.174\pm0.027~\mathrm{Jy~km~s^{-1}}$ ([\Cii]1030A) and $0.276\pm0.044~\mathrm{Jy~km~s^{-1}}$ (B), respectively, with the FWHM line width $64\pm11~\mathrm{km~s^{-1}}$ ([\Cii]1030A) and $62\pm11~\mathrm{km~s^{-1}}$ (B).
These fluxes are the values after correction for the primary beam response (67\% and 47\% at the positions of [\Cii]1030A and B, respectively), and for the $\approx20\%$ loss falling outside the aperture of $0.24\arcsec$ in radius for extracting the spectra.  

The corresponding [\Cii] luminosity is $L_\textrm{[C\,{\sc ii}]} = (1.61\pm0.24)\times10^8~{L_\odot}$ and $(2.21\pm0.36)\times10^8~{L_\odot}$, respectively, for [\Cii]1030A and B.  An empirical conversion relation of $\mathrm{SFR=-7.06+1.0\log_{10}(L_\textrm{[C\,{\sc ii}]}}/L_\odot)$ yields a [\Cii]-based SFR of $\mathrm{SFR_\textrm{[C\,{\sc ii}]}} = 14~M_\odot~\mathrm{yr^{-1}}$ ([\Cii]1030A) and $19~M_\odot~\mathrm{yr^{-1}}$ (B), respectively, with 0.3~dex uncertainty  \citep{2014A&A...568A..62D}.  Although more recent studies inferred closely consistent conversion relations using high-redshift [\Cii] sources \citep{2020A&A...643A...3S}, the [\Cii] luminosity at fixed SFR is likely to drop when the metallicity is low ($Z<0.1\textrm{--}0.2~Z_\odot$; \citep{2015ApJ...813...36V,2017ApJ...836L...2B}).  For a lower metallicity, the [\Cii]-based SFR would be even higher and thus requires even more dust obscuration, which is in turn at odds with the low metallicity.  Therefore, these [\Cii] sources are likely to be as metal rich as the levels of normal [\Cii] sources detected at similar redshifts \citep{2020A&A...643A...3S,2022MNRAS.512...58F}.

\begin{figure}[t]
\centering
\includegraphics[width=4.0in]{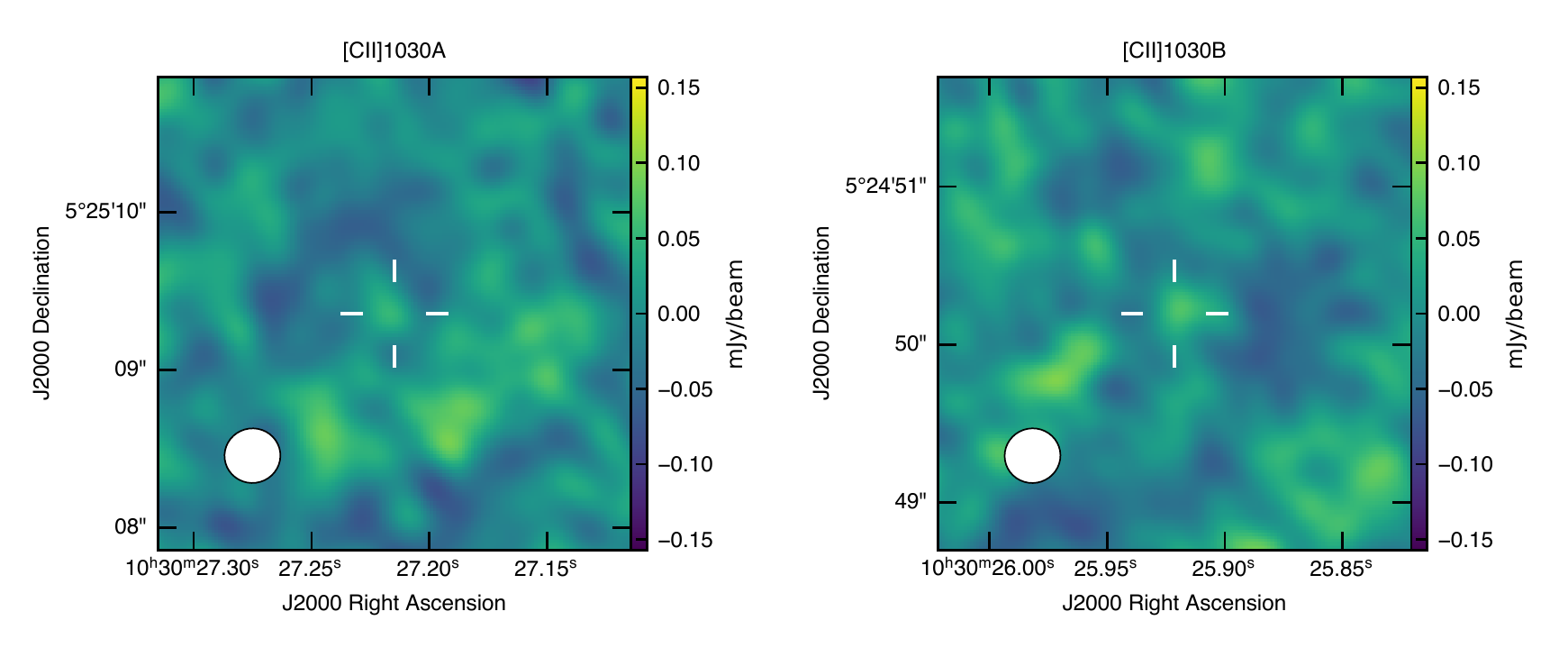}
\caption{\label{fig:continuum_maps}
\textbf{Continuum maps.}  Each image cutouts the $3\arcsec\times3\arcsec$ region centered at the position of [\Cii]1030A (left panel) and B (right panel).  The rest-frame 1900~GHz continuum intensity is shown by the color scale.  The synthesized continuum beam size is indicated by the white circle in the lower left corner.}
\end{figure}

The ALMA datacubes cover the rest-frame FIR continuum at 1900~GHz that traces the dust emission.  The continuum maps of the sources are shown in Extended Data Fig.~\ref{fig:continuum_maps}.
We found no detection of the continuum at the positions of the [\Cii] sources and inferred a $2\sigma$ upper limit of $106~\mathrm{\mu Jy}$ ([\Cii]1030A) and $158~\mathrm{\mu Jy}$ (B).
To convert this continuum flux to a total infrared luminosity ($L_\mathrm{IR}$; $8\textrm{--}1000~\mathrm{\mu}m$), we adopted a conversion factor of $\nu L_\nu/L_\mathrm{IR}=0.13$ ($\nu=1900~\mathrm{GHz}$) that was inferred using a template of the infrared spectral energy distribution (SED) of $z\sim4$ galaxies \citep{2020A&A...643A...2B}.  The template is consistent with a modified blackbody ($S_\nu \approx \nu^{\beta}/(\exp(h\nu/kT_\mathrm{dust})-1)$) with a fixed index $\beta=1.8$ and dust temperature $T_\mathrm{dust}=41~\mathrm{K}$.  This conversion resulted in a $2\sigma$ upper limit of $L_\mathrm{IR}=2.3\times10^{11}~L_\odot$ (A) and $3.3\times10^{11}~L_\odot$ (B).  Finally, we converted the $L_\mathrm{IR}$ to obscured SFR by adopting the relation ($\mathrm{SFR}_\mathrm{IR}(M_\odot~\mathrm{yr^{-1}})=1.01\times10^{-10}~L_\mathrm{IR}(L_\odot)$) \citep{2014ARA&A..52..415M} converted to a \citep{2003PASP..115..763C} initial mass function, and obtained $\mathrm{SFR}_\mathrm{IR}<23~M_\odot~\mathrm{yr^{-1}}$ (A) and $<34~M_\odot~\mathrm{yr^{-1}}$ (B) at $2\sigma$.  The $2\sigma$ lower limits of $L_\mathrm{[CII]}/L_\mathrm{IR}$ ratio, $>7 \times10^{-4}$, is consistent with the observed values at similar redshifts \citep{2020A&A...643A...3S} with no evidence of the so-called [CII]-deficit.

\subsection*{Measurements of the absorption systems toward QSO J1030$+$0524}

Our analysis uses revised measurements of the absorption systems at $z \approx 5.7$ toward QSO J1030+0524, using improved spectra obtained with both VLT/XShooter \citep{2011A&A...536A.105V} and the Folded-port InfraRed Echellette (FIRE) on Magellan \citep{2013PASP..125..270S}.  
Archival X-shooter data (acquired under ESO programme 084.A-0360(A), 086.A-0162(A), 086.A-0574(A), and 087.A-0607(A)) were retrieved from the ESO Science Archive Facility and reduced using the PypeIt spectroscopic reduction pipeline \citep{2020JOSS....5.2308P}. 
The XShooter exposures from different programs were carefully and individually corrected for telluric absorption using numerical atmospheric models and then optimally combined to maximize signal-to-noise ratio.
All exposures were obtained with a $0.9\arcsec$ slit, resulting in spectral resolution is $\lambda/\Delta \lambda = 5300$. 
FIRE expoures were acquired in April 2011 and reduced using the FIREHOSE pipeline \citep{jonathan_gagne_2015_18775}. 
All data used a $0.6\arcsec$ slit for spectral resolution $\lambda/\Delta \lambda = 6000$. 
The FIRE spectra were corrected for telluric absorption using observations of A0V standards and a modified version of the xtellcor package \citep{2003PASP..115..389V,2004PASP..116..362C}

Because the absorption data span three instrument configurations (XShooter Visible and NIR channels and FIRE), we have developed absorption-line fitting routines that generate a single hierarchical physical model of components and transitions, and then project that model onto each instrument's respective spectral data, convolving to each spectrogaph's appropriate resolution. 
The model vector is optimized to fit all instruments simultaneously using a Markov-Chain Monte Carlo walker as implemented by the emcee python package \citep{2013PASP..125..306F}, to determine confidence intervals from the posterior distribution. 
Prior to fitting absorption models, each spectrum was normalized by a continuum iteratively determined by cubic spline fits with outlier rejection.

\begin{figure}[t]
\centering
\includegraphics[width=4.0in]{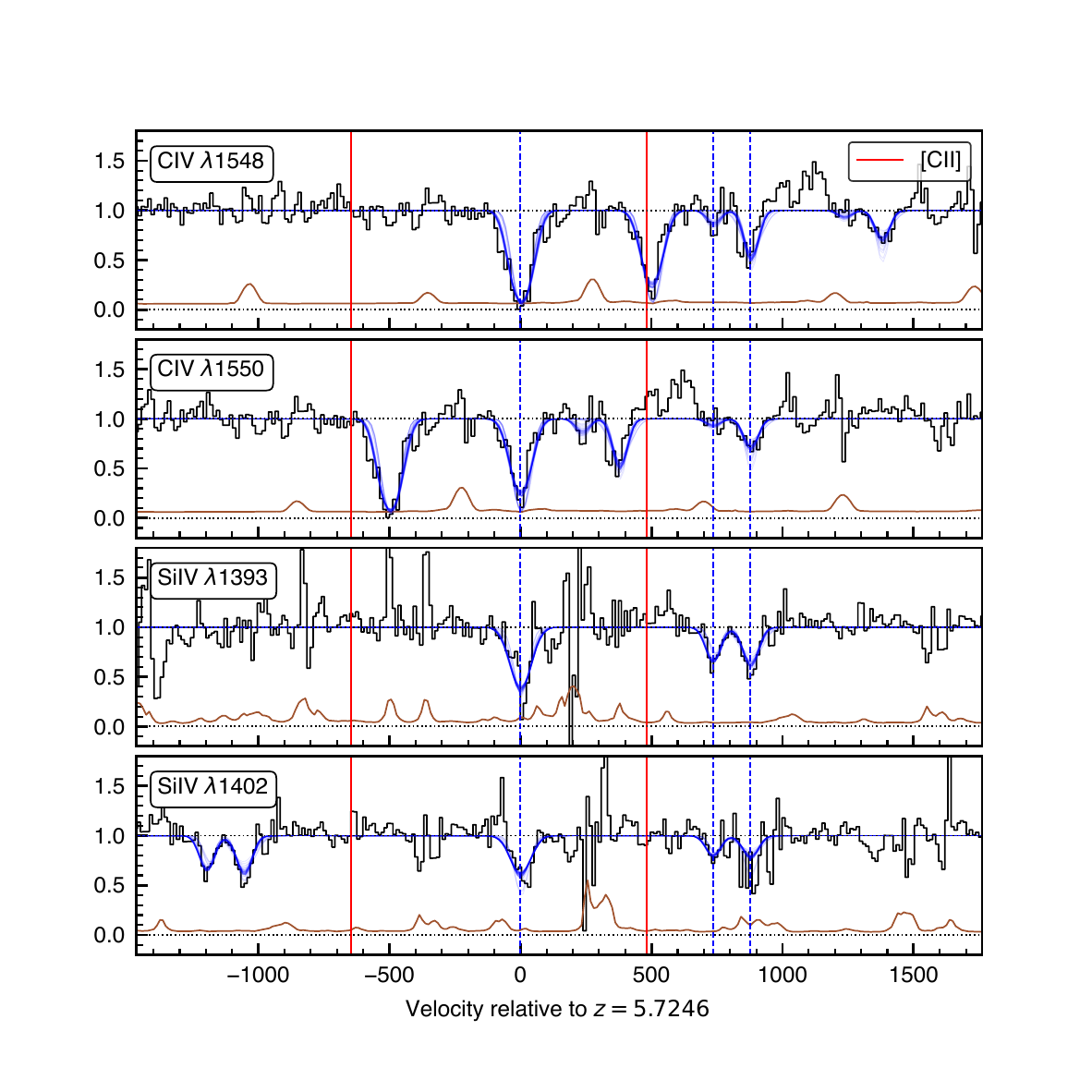}
\caption{\label{fig:voigt}
\textbf{{\Civ} and {\Siiv} absorption systems at $z_\mathrm{abs}\approx5.7$.}  Each panel corresponds to the transition labeled in the upper left.  The black line shows the observed spectrum of QSO J1030 while the light brown line denotes the standard deviation of the flux.  The blue lines indicate 50 random Voigt profiles sampled from the posterior distribution.  The origin of the velocity scale was chosen to correspond to the strongest absorption system at $z_\mathrm{abs}=5.7246$, which is marked with the two other systems ($z_\mathrm{abs}=5.7411$ and 5.7443) in the blue dashed lines.  The red vertical lines indicate the relative velocities of [\Cii]1030A and B.}
\end{figure}

For this particular absorption complex, we identified three distinct components of high-ionization gas with {\Civ} and {\Siiv} aligned in velocity space.  
The adopted Voigt profile model constrains redshifts of the {\Civ} and {\Siiv} components to align in velocity.  As seen in Extended Data Fig.~\ref{fig:voigt}, the {\Civ} component at $z_\mathrm{abs}=5.7411$ is a marginal identification with low statistical significance. 
However, a firm {\Siiv} detection is seen for all three components.  Numerous low-ionization species are also present at $z\sim 5.7$; these are not shown and we defer them to a later analysis as some low-ionization transitions are blended with lower-redshift interloping absorbers.

Table \ref{tb:Voigt} lists the Voigt profile median and [16\%, 84\%] confidence intervals on the fit. It should be noted that XShooter and FIRE likely do not resolve fine velocity structures, so the Doppler parameters $b$ in the fits may represent a velocity dispersion among components that would be separated at higher resolution, rather than the intrinsic width of a single component.  However, tests indicate that the total column densities of these fits are robust when interpreted as a sum of the unresolved blend.

\subsection*{HST broadband imaging}

In the field of QSO J1030$+$0524, we carried out HST broadband imaging in the F775W (mean wavelength $\lambda_\mathrm{mean}=7731${\AA}) and F160W ($\lambda_\mathrm{mean}=15436${\AA}) bands (program IDs 13303 and 15085).  In the HST archival database, additional observations (program ID 9777) in F775W and F850LP ($\lambda_\mathrm{mean}=9080${\AA}) are available.  All the HST images were processed and co-added using the DrizzlePac software \cite{drizzlepac2}.

\begin{figure}[t]
\centering
\includegraphics[width=4.0in]{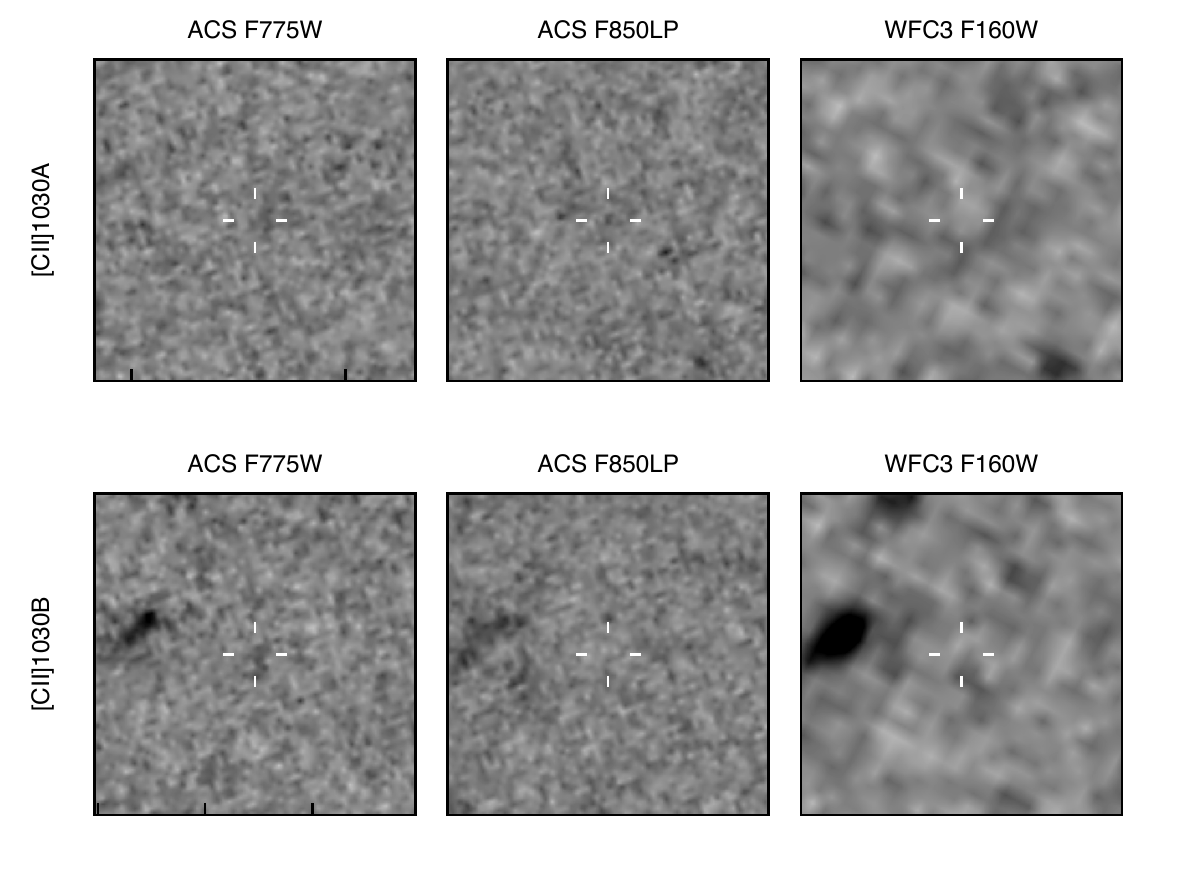}
\caption{\label{fig:HST_cutouts}
\textbf{HST broadband images.}  Each image cutouts the $3\arcsec \times 3\arcsec$ region centered at the position of the [\Cii]1030A (upper row) and B (lower row).  From left to right, these images are F775W, F850LP, and F160W, respectively, which sample the rest-frame 1150~{\AA}, 1350~{\AA}, and 2300~{\AA}.  No significant rest-frame UV counter part is detected at the positions of the [\Cii] emission.}
\end{figure}

These images were used to search for rest-frame ultraviolet counterparts of the claimed [\Cii] emitters.  For sources at $z\approx5.7$, the F775W, F850LP, F160W bands sample the rest-frame 1150~{\AA} (including the wavelength of Lyman-$\alpha$), 1350~{\AA}, and 2300~{\AA}, respectively.  The $3\arcsec\times3\arcsec$ cutout images are shown in Extended Data Fig.~\ref{fig:HST_cutouts}.  No significant counterpart was found at the positions of the [\Cii] sources, yielding $2\sigma$ upper limits of 45~nJy (F775W), 59~nJy (F850LP), and 31~nJy (F160W) for the aperture photometry in $0.7\arcsec$-diameter.
Here the standard deviation of the photometric flux is estimated by conducting aperture photometry at random sky positions.

The rest-frame FUV flux traces unobscured star formation of a galaxy.  
We converted the F850LP fluxes to the UV-based $\mathrm{SFR_{UV}}$ adopting a relation ($\mathrm{SFR}_\mathrm{FUV}~(M_\odot~\mathrm{yr^{-1}})=0.72\times10^{-28} L_\nu (\mathrm{erg~s^{-1}~Hz^{-1}})$) \citep{2014ARA&A..52..415M}, yielding $\mathrm{SFR}_\mathrm{UV}<2.4~M_\odot~\mathrm{yr^{-1}}$.

\subsection*{MUSE observation}

We retrieved the archival data of the integral field spectroscopy in this field taken with the MUSE on the VLT UT4 under ESO programme 095.A-0714(A).  The datacube covers a $1\arcmin \times 1\arcmin$ region centered at the position of QSO J1030$+$0524 across the wavelength range of 4750--9350~{\AA} with the total on-source exposure time of 6.4 hrs. The observations were carried out in the WFM-NOAO-N mode in excellent seeing. The spatial FWHM is $\simeq0.6''$ measured on the ``white light'' image, constructed from collapsing the reduced datacube over the full wavelength range.

The initial data reduction was carried out with the MUSE pipeline (v2.8.5, \citep{2020A&A...641A..28W}). 
With the pipeline we applied bias subtraction, flat fielding, wavelength calibration and produced datacubes and pixel tables. The quality of the sky subtraction in the pipeline reduced version is limited by flat fielding errors, due to temperature fluctuations in the instrument. As a post-processing step, we apply CubEx (see e.g. \citep{2016ApJ...831...39B,2019MNRAS.483.5188C}) to perform self-calibration to improve the flat fielding and sky subtraction.  MUSE datacubes often contain small shifts ($\lt 2\arcsec$) in the field center with respect to the astrometry solution in the header, this is caused by the so-called ``derotator wobble'' \citep{2015A&A...575A..75B}.
In order to facilitate matched spectroscopy for the ALMA detections in MUSE it was essential that the astrometry be well calibrated. 
The MUSE field of view does not contain any stars with Gaia astrometry, so instead we used the HST/ACS F755W image as a bridge to propagate Gaia calibration to the MUSE. 

We used the MUSE datacube to search for counterparts of [\Cii]1030A and B in the Ly$\alpha$ emission line within search radii of $\approx 3\arcsec$, or 20~pkpc, in position and $\pm 500~\mathrm{km~s^{-1}}$) in velocity.  No significant Ly$\alpha$ emission was detected at the expected positions and wavelengths for these [\Cii] sources, yielding a 2$\sigma$ upper limit of $\approx 1\times10^{-18}~\mathrm{erg~s^{-1}~cm^{-2}}$ on the observed Ly$\alpha$ fluxes.

\bmhead{Data availability}

The ALMA, HST, and VLT data sets used in this work are publicly available, respectively, through the ALMA data archive at \url{https://almascience.nao.ac.jp/aq/} (project code 2017.1.00621.S), the HST data archive at \url{https://archive.stsci.edu/missions-and-data/hst} (Proposal IDs 9777, 13303, and 15085), and the European Southern Observatory Science Archive at \url{http://archive.eso.org} (Programme ID 084.A-0360, 086.A-0162, 086.A-0574, and 087.A-0607, and 095.A-0714).

\bmhead{Code availability}

The ALMA data was processed using the CASA pipeline version 5.4, which is available at \url{https://casa.nrao.edu}.  
The  \texttt{interferopy} package that includes the \texttt{findclumps} source finding program is available at \url{https://github.com/interferopy/interferopy}.  
The VLT/X-shooter spectral data were reduced using the PypeIt package available at \url{https://github.com/pypeit/PypeIt}.
The Magellan FIRE spectral data ware reduced using the FIREHOSE pipeline available at \url{https://wikis.mit.edu/confluence/display/FIRE/FIRE+Data+Reduction} and corrected using the xtellcor package available at \url{http://irtfweb.ifa.hawaii.edu/~spex/index.html}.
The DrizzlePac software used for the HST data reduction is available at \url{https://www.stsci.edu/scientific-community/software/drizzlepac.html}.
The reduction pipelines of the VLT instruments are available at \url{https://www.eso.org/sci/software/pipelines/}.

\bmhead{Acknowledgments}

This work is based on the ALMA observation under ADS/JAO.ALMA\#2011.0.01234.S.  
ALMA is a partnership of ESO (representing its member states), NSF (USA) and NINS (Japan), together with NRC (Canada), MOST and ASIAA (Taiwan), and KASI (Republic of Korea), in cooperation with the Republic of Chile. 
The Joint ALMA Observatory is operated by ESO, AUI/NRAO and NAOJ.  
This research makes use of observations made with the NASA/ESA Hubble Space Telescope obtained from the Space Telescope Science Institute, which is operated by the Association of Universities for Research in Astronomy, Inc., under NASA contract NAS 5–26555. This paper includes data gathered with the 6.5 meter Magellan Telescopes located at Las Campanas Observatory, Chile.  Data analysis was in part carried out on the Multi-wavelength Data Analysis System operated by the Astronomy Data Center, NAOJ.
We acknowledge support from JSPS KAKENHI Grant Number JP21K13956. 

\bmhead{Author Contribution}

DK, SJL, RB, RAS discussed and planned the ALMA program and the HST observations in the F775W band. 
RAS obtained the HST observation in the F160W band.
ACE and RAS analyzed the X-shooter and FIRE spectroscopic data of QSO~J1030$+$0524.
RM reduced and calibrated the MUSE data.
SJL, JM, RAS, RM, and ACE contributed to the discussion of the presented results and the prepareation of the manuscript.
DK led the team, being principal investigator of the ALMA program, analyzed the ALMA and ancillary data, wrote the main text and the Methods section, and produced all figures and the table in the article.

\bmhead{Conflict of interest/Competing interests}

The authors declare no competing interests.

\clearpage

\bigskip
\bigskip

\newpage~\newpage~\newpage~\newpage~\newpage~\newpage
\begin{table}[h]
\begin{center}
\begin{minipage}{\textwidth}
\caption{\textbf{Expected number counts of CO sources.} The frequency is given in the rest-frame.  The redshift and luminosity are calculated under the assumption that the fainter line (A) detected in the ALMA data is each of the CO transition.  The expected number counts of unrelated CO sources at a line flux brighter or equal to the observed flux are provided in the fourth column.}\label{tb:CO_interlopers}%
\begin{tabular}{@{}ccccc@{}}
\toprule
Transition & frequency & Redshift & $\log L_\mathrm{CO}$            & $N$ \\
           & (GHz)     &          & ($\mathrm{Jy~km~s^{-1}~Mpc^2}$) & \\
\midrule
$J=3$--2 & 345.80 & 0.22 & 6.46 & 0.02 \\
$J=4$--3 & 461.04 & 0.63 & 7.51 & 0.11 \\
$J=5$--4 & 576.27 & 1.04 & 8.04 & 0.11 \\
$J=6$--5 & 691.47 & 1.45 & 8.40 & 0.10 \\
$J=7$--6 & 806.65 & 1.85 & 8.66 & 0.05 \\
$J=8$--7 & 921.80 & 2.26 & 8.87 & 0.02 \\
\midrule
Total & & & & 0.40 \\
\botrule
\end{tabular}
\end{minipage}
\end{center}
\end{table}

\begin{table}[h]
\begin{center}
\begin{minipage}{\textwidth}
\caption{\textbf{Absorption-line measurements.}  The median and [16\%, 84\%] confidence intervals are given for each of the Voigt profile parameters.  The second and third columns present the column densities of {\Civ} and {\Siiv}.  The fourth column is the Dopper parameter $b$ } \label{tb:Voigt}%
\begin{tabular}{@{}ccccc@{}}
\toprule
Redshift & $\log N(\textrm{\Civ})$\footnotemark[2] & $\log N(\textrm{\Siiv})$\footnotemark[3] & $b$\footnotemark[4] \\
         & ($\mathrm{cm}^{-2}$)    & ($\mathrm{cm}^{-2}$)     & ($\mathrm{km~s^{-1}}$) \\
\midrule
5.72457 [5.72455, 5.72460]\footnotemark[1] & 14.55 [14.52, 14.63] & 13.70 [13.68, 13.72] & 40.6 [35.0, 42.6] \\
5.74106 [5.74102, 5.74110] & 13.08 [12.68, 13.20] & 13.26 [13.23, 13.29] & 20.8 [7.0, 22.6] \\
5.74426 [5.74422, 5.74428] & 13.81 [13.77, 13.85] & 13.31 [13.28, 13.35] & 23.0 [20.6, 30.6] \\
\botrule
\end{tabular}
\footnotetext[1]{The median and [16\%, 84\%] confidence intervals are given for each of the Voigt profile parameters.}
\footnotetext[2]{The column density of {\Civ}.}
\footnotetext[3]{The column density of {\Siiv}.}
\footnotetext[4]{The Doppler parameter $b$ that is related to the kinetic temperature $T$ of the gas via $b=\sqrt{2kT/m_\mathrm{p}}$ where $k$ is the Boltzmann constant and $m_\mathrm{p}$ is the proton mass.}
\end{minipage}
\end{center}
\end{table}

\backmatter





\end{document}